\documentstyle[epsfig,epsf]{article}
\def\nmne {\nu_\mu \longleftrightarrow \nu_e}
\def\nmnt {\nu_\mu \longleftrightarrow \nu_\tau}

\def\ne {\nu_e}
\def\nm {\nu_\mu}
\def\nt {\nu_\tau}
\def\dm2 {\Delta m^2}
\def\s2t {\sin^2 2\theta}

\begin{document}

\begin{center}
{\Large {\bf The CNGS Neutrino Beam}}
\end{center}

\vskip .7 cm

\begin{center}
G. GIACOMELLI  \par
{\it Dept of Physics, Univ. of Bologna and INFN, \\
V.le C. Berti Pichat 6/2, Bologna, I-40127, Italy\\} 

E-mail: giacomelli@bo.infn.it

\par~\par

Lecture given at the 2$^{nd}$ Latin American School on Cosmic Rays and 
Astrophysics, Puebla, Mexico,\\ 30$^{th}$ August - 8$^{th}$ September 2006.

\vskip .7 cm
{\large \bf Abstract}\par
\end{center}

{\normalsize 
The CERN to Gran Sasso Neutrino beam (CNGS) was commissioned at CERN in early 
August 2006 and was first sent at low intensity to Gran Sasso on August 17, 
2006. The Borexino, LVD and OPERA detectors continued the commissioning of 
their detectors and started taking data with practically no dead time. The CNGS operated smoothly with good quality. In a short time the 3 detectors 
collected several hundred events with clean time distributions.
}


\large
\section{Introduction}\label{sec:intro}
Neutrino physics has opened new windows into phenomena beyond the Standard 
Model of particle physics. Long baseline neutrino experiments may allow
further insight into neutrino physics. The CERN to Gran Sasso neutrino beam 
(CNGS) is one of these projects \cite{CNGS}; it was commissioned at CERN in 
early august 2006 and it started sending beam to the Gran Sasso Lab. (LNGS) 
on the 17$^{th}$ of August 2006. At Gran Sasso 3 detectors, Borexino, LVD and OPERA, were ready to use it, and imediately started to see beam neutrino 
events. \par
OPERA \cite{opera} is a hybrid-emulsion-electronic detector,
 designed to search 
for the $\nmnt$ oscillations in the 
parameter region indicated by the MACRO \cite{macrobib}, SuperKamiokande 
\cite{skbib} and Soudan2 \cite{soudanbib} 
atmospheric neutrino results \cite{gg-io}, recently confirmed
by the K2K \cite{k2k} and MINOS \cite{minos} long baseline experiments.
 One of the main goals of OPERA is to find the $\nt$ appearance by direct 
detection of the $\tau$ lepton from $\nt$ CC interactions. 
The detection of the 
$\nt$ will be made via the charged 
$\tau$ lepton produced in $\nt$ CC interactions, and its decay products. 
To observe the decays, a spatial resolution of 
$\sim 1~\mu$m is necessary; this resolution is obtained in emulsion 
sheets interspersed with thin lead target plates. This technique, the 
``Emulsion Cloud Chamber'' (ECC), was started in the $\tau$ search 
experiment \cite{donutbib}. OPERA may 
also search for the subleading $\nmne$ oscillations and make a 
variety of observations with or without the beam using its electronic 
subdetectors. \par
LVD is an array of 840 liquid scintillators with a total mass of 
1000 t; it is designed to search and study neutrinos from gravitational stellar collapses \cite{www}. LVD plans to be a neutrino flux monitor of the CNGS 
beam. \par
Borexino is a refined electronic detector designed to study solar neutrinos, 
in particular the monoenergetic neutrinos coming from Be$^7$ 
decay \cite{http}. In August 2006 it was only partially filled with water: now 
it is completely filled with water and its inner part is partially filled with 
liquid scintillator.  \par
In this lecture note will be summarized and discussed the CNGS neutrino beam, some features of the Borexino, LVD, and OPERA experiments, and the preliminary results obtained by the three experiments during the August 2006 test run.

\begin{figure}[!ht]
\begin{center}
\mbox{\epsfig{figure=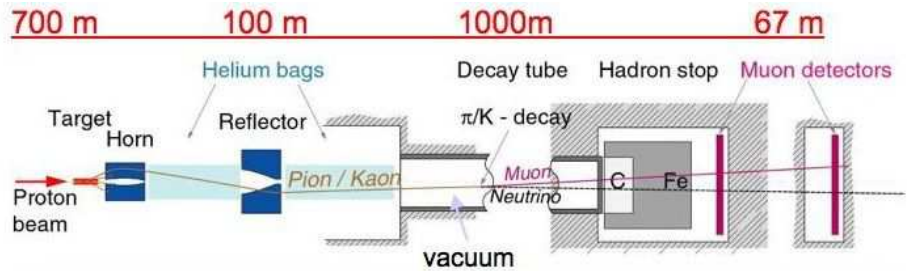,height=4.5cm}}
\mbox{\epsfig{figure=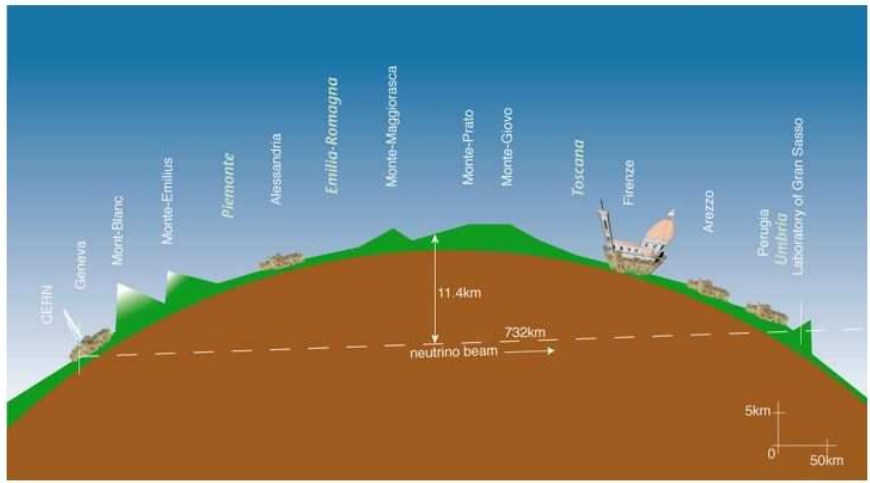, height=8.5cm}}
\caption {Top: The main components of the CNGS neutrino beam at CERN. 
Bottom: Sketch of the 730 km neutrino trajectory from CERN to Gran Sasso.}
\label{fig:CNGS}
\end{center}
\end{figure}

\section{The CNGS neutrino beam}
\label{sec:CNGS}
Fig. \ref{fig:CNGS} top shows the main components of the $\nm$ beam 
at CERN \cite{CNGS}. A 400 GeV proton beam is extracted from the SPS and is 
transported to the CNGS target. Secondary pions and kaons of positive charge 
are focused into a parallel beam by two magnetic lenses, called horn  and 
reflector. A long decay pipe allows the pions and 
kaons to decay 
into $\nu_{\mu}$  and $\mu$. The remaining hadrons are absorbed in a beam 
dump. The muons are monitored by two sets of detectors downstream of the dump.

\begin{figure}[!ht]
\begin{center}
\mbox{\epsfig{figure=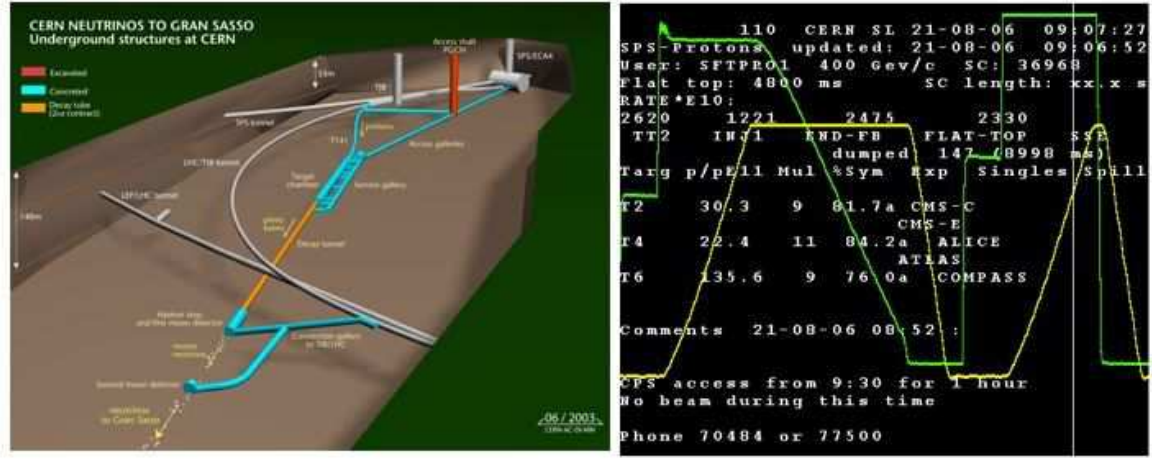,height=7.5cm}}
\caption {Left: Underground layout of the SPS and of the CNGS beam at CERN. 
Right: Scheme of the SPS operation at CERN during the August test run.}
\label{fig:CNGSbeam}
\end{center}
\end{figure}

\begin{figure}[!ht]
\begin{center}
\mbox{\epsfig{figure=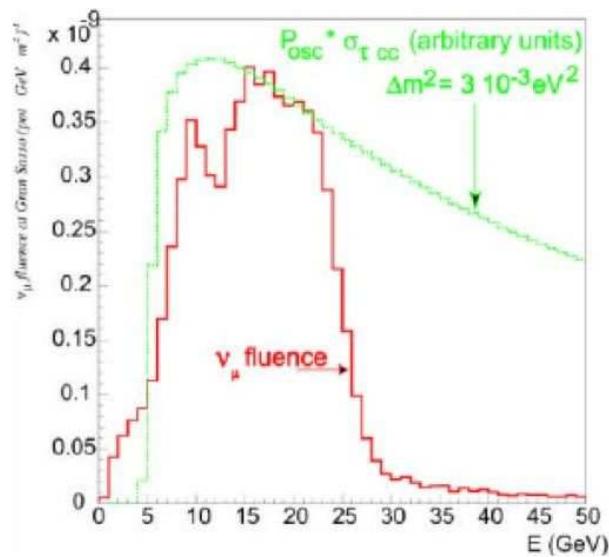,height=7.9cm}}
\caption {Energy distribution of the CNGS $\nm$ beam at Gran Sasso.}
\label{fig:spettro}
\end{center}
\end{figure}

Fig. \ref{fig:CNGS} bottom shows the 730 km path of the 
CNGS neutrinos from CERN to Gran Sasso. The beam is optimised for producing 
a maximum number of CC $\nt$ interactions in OPERA at Gran Sasso. 
Fig. \ref{fig:CNGSbeam} left shows the underground layout of the SPS and of CNGS at CERN; Fig. \ref{fig:CNGSbeam} right shows the scheme of the SPS operation 
during the August 2006 test run.
The energy distribution of the beam 
at Gran Sasso is shown in Fig. \ref{fig:spettro}: the mean $\nm$ beam energy is 17 GeV, the $\bar{\nm}$ 
contamination is $\sim$2$\%$, the $\ne$ ($\bar{\ne}$) is $< 1\%$ and the 
number of $\nt$ is negligible. The $L/E_\nu$ ratio is 43 km/GeV. 
The muon beam size at the second muon detector at CERN is about 
$\sigma \sim$1 m; this translates into a neutrino beam size at GS of 
$\sigma \sim$1 km. Civil 
engineering was completed in june 2006, all beam parts were installed and 
commissioning was made in early August. The first low intensity test beam was 
expected at GS in August 17 2006: this happened, and almost immediately the 3 
detectors at GS obtained their first events (Figs. \ref{fig:display}, 
\ref{fig:display4}). 
The low intensity CNGS was stable 
and of high quality. The shared SPS beam sent a pulse of 2 neutrino bursts,
each of 10.5 
$\mu$sec duration, separated by 50 ms, every 12 s \cite{CNGS}. 
A higher intensity beam was expected 
for October 2006, but it did not happen because of a water leak in the beam at 
CERN.

\section{BOREXINO}
\begin{figure}[!ht]
\begin{center}
\mbox{\epsfig{figure=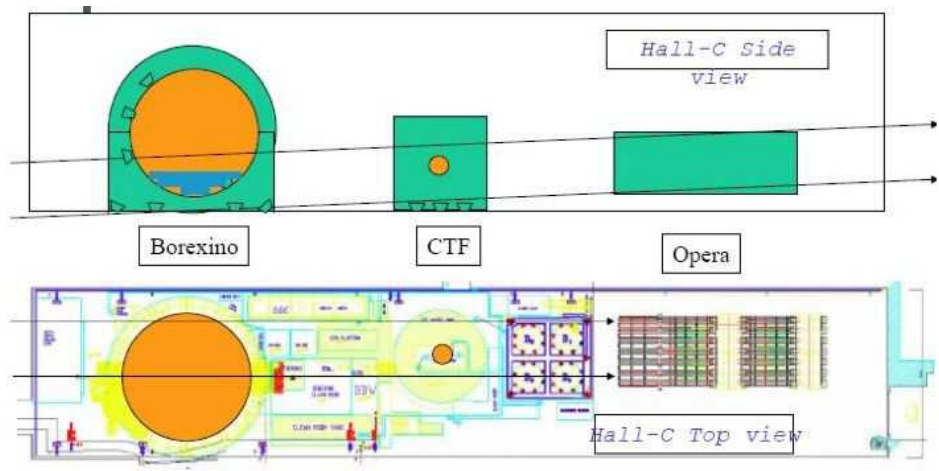,height=6.8cm}}
\caption {Layout of the Borexino and OPERA experiments in Hall C of the 
underground Gran Sasso Lab. Side view of Hall C (upper figure) 
and view from the top (lower figure).}
\label{fig:hallC}
\end{center}
\end{figure}

The main aim of the BOREXINO detector is the measurement of the monochromatic 
$\nu_e$'s coming from Be$^7$ decays in the center of the sun. The layout of 
BOREXINO in Hall
 C is shown in Fig. \ref{fig:hallC} \cite{http}. The most important part of the detector is 
the large sphere, whose  central part should be filled with liquid 
scintillator. The 
outer sphere should be filled with pure water. The filling with water was started in early August and during the first run only few meters of water were in the sphere. The partial filling limited the data acquisition. Nevertheless 5 
neutrino candidate events 
were observed. In December 2006 the water filling was completed and initial 
filling with liquid scintillator in the inner sphere was started.  

\section{LVD}
\label{sec:LVD}
\begin{figure}[!ht]
\begin{center}
\mbox{\epsfig{figure=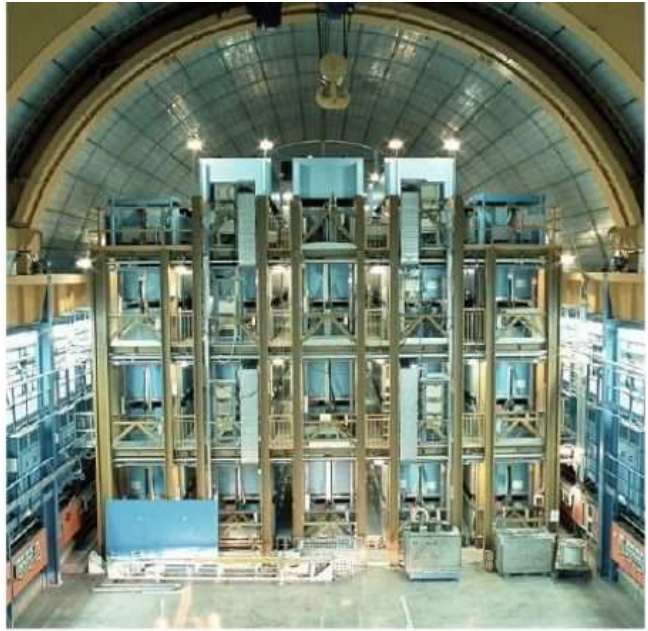,height=8.5cm}}
\caption {The LVD detector in the Gran Sasso Hall A.}
\label{fig:LVD}
\end{center}
\end{figure}

The main purpose of the LVD detector is the search for electron antineutrinos 
from gravitational stellar 
collapses in our galaxy. LVD is made of three identical ``towers'', 
each containing 8 active modules; a module has 8 counters of dimension 
$1 \times 1 \times 1.5~ m^3$, filled with 1.2 t of liquid scintillator. 
Fig. \ref{fig:LVD} shows the LVD detector located in Hall A of Gran Sasso; 
it has a total 
mass of 1000 t \cite{www}. Neutrinos from the CNGS beam are observed through: (i) the detection of muons produced in neutrino CC interactions in the surrounding rock or in the detector, (ii) through the detection of the hadrons produced in neutrino 
NC/CC interactions inside the detector.\par

\begin{figure}[!ht]
\begin{center}
\mbox{\epsfig{figure=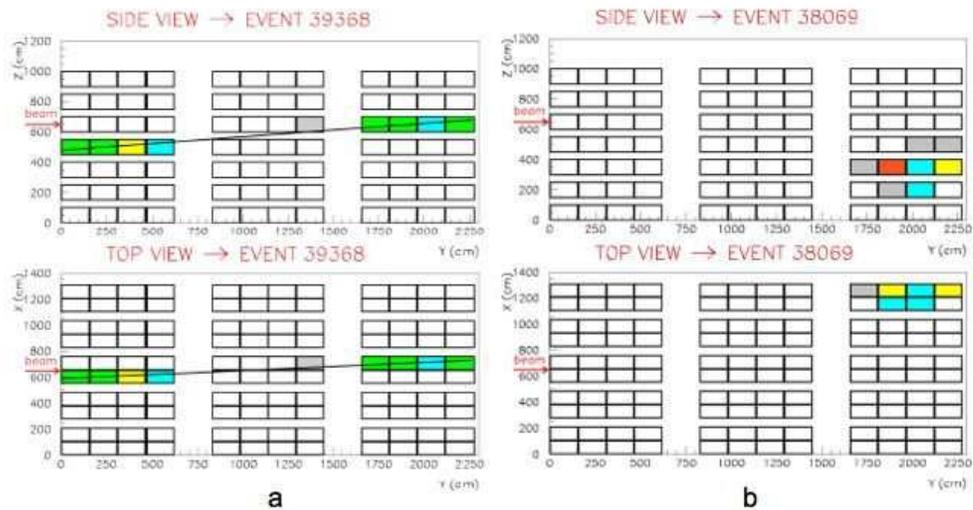,height=8.5cm}}
\caption {Observation in LVD of (a) one muon originated in a $\nu_{\mu}$ CC 
interaction in the 
rock before the detector; (b) a multihadron event originated by a $\nu_{\mu}$ 
NC/CC interaction inside the detector.}
\label{fig:display}
\end{center}
\end{figure}

LVD plans to be a neutrino flux monitor of CNGS at Gran Sasso. Since LVD is 
running 
since several years, it was completely ready when the neutrino beam arrived. 
Fig. 6a shows a muon coming from a $\nu_{\mu}$ interaction in the rock before the detector; Fig. 6b shows a $\nu_{\mu}$ NC/CC interaction inside the detector. 
In the August 2006 test run LVD was counting 50-100 muons per day and  
recorded a total of about 500 events \cite{www}.

\section{OPERA}
\label{sec:construction}
\begin{figure}[!ht]
\begin{center}
\mbox{\epsfig{figure=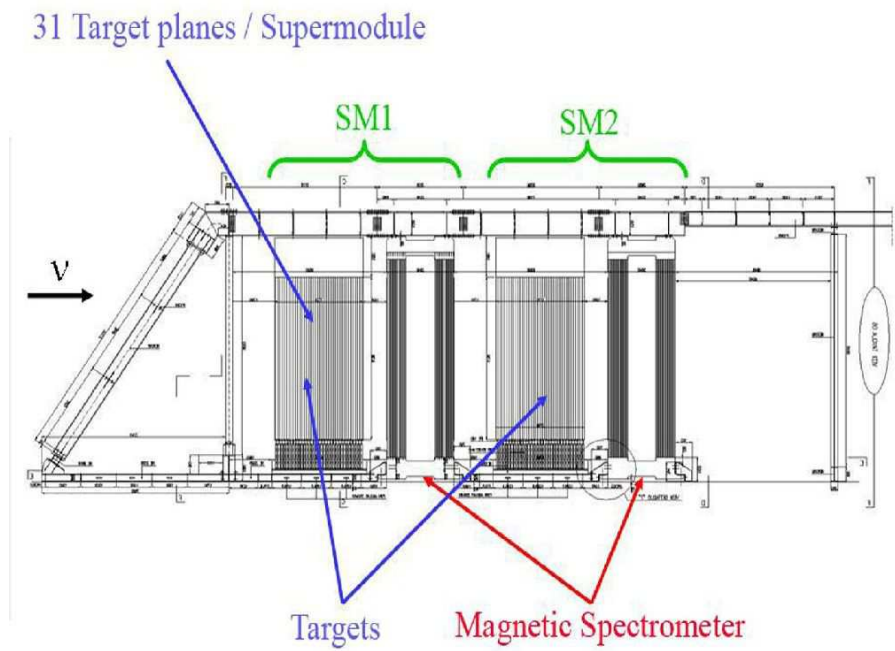,height=8.5cm}}
\caption {Layout of the OPERA detector.}
\label{fig:detopera}
\end{center}
\end{figure}

The OPERA detector, Fig. \ref{fig:detopera}, is a hybrid detector made of two 
identical supermodules, each consisting of a target section 
with 31 target planes of lead/emulsion-film modules (``bricks''), of a scintillator tracker detector and of a muon spectrometer. The initial target 
mass should be 1.8 kt. It is the first detector 
specifically designed for the CNGS $\nu$ beam.

{\bf Electronic subdetectors.} The first electronic subdetector is an 
{\it anticoincidence wall} to better separate muon events coming from 
interactions in OPERA from those in the material and rock  before OPERA.\par
The {\it target tracker} is made of 32000 scintillator strips, each 7 m 
long and of 25 mm$ \times $15 mm cross section (7000 m$^2$ area). Along 
the strip, a wavelength shifting 
fibre of 1 mm diameter transmits the light signals to both ends. The readout 
is done by 1000 64 channel Hamamatsu PhotoMultipliers (PMTs). \par
The {\it muon spectrometer} consists of 2 iron magnets instrumented with 
{\it Resistive Plate Chambers} (RPC) and {\it drift tubes}. Each magnet is an 
$8 \times 8$ m$^2$ dipole with a field 
of 1.52 T in the upward direction on one side and in the downward direction on 
the other side. This allows to measure the momentum twice, 
 reducing the error by $\sqrt{2}$. A magnet consists of 
twelve 5 cm thick iron slabs, alternated with RPC planes. In the magnetic 
field a muon is tracked, identified and its momentum is measured.\par
The {\it precision tracker} 
\cite{specbib} measures the muon track coordinates in the horizontal 
plane. It is made of 12 drift tube planes, each covering an area of 
$8 \times 8$ m$^2$; they are placed in front and behind each 
magnet and between the two magnets. 
The muon spectrometer allows a momentum resolution 
$\Delta p / p \le 0.25$ for muon momenta $< 25$ GeV/c. 
 Two planes of {\it glass 
RPC's (XPC's)}, consisting of two $45^{\circ}$ crossed planes, 
 are installed in front of the magnets.\par
In August 2006 the brick supporting structure, the tracker planes, 
 the XPC's and three of the high
precision tracker planes of the first supermodule were installed. The 
magnets, including all RPC's and the mechanical structure were completed. \par
The DAQ system uses a 
Gigabit network of 1150 nodes. To match the data of the different 
subdetectors an event ``time stamp'' is delivered by a clock using the Global 
Positioning System (GPS). The synchronization with the beam spill is done via 
GPS. 
The DAQ uses a system which contains 
the CPU, the memory, the clock receiver for the time stamp and the ethernet 
connections to the other components. \par
The commissioning of each electronic subdetector was made first with cosmic 
ray muons and then with the CNGS at reduced intensity in August 2006.\par

\begin{figure}[!ht]
\begin{center}
\mbox{\epsfig{figure=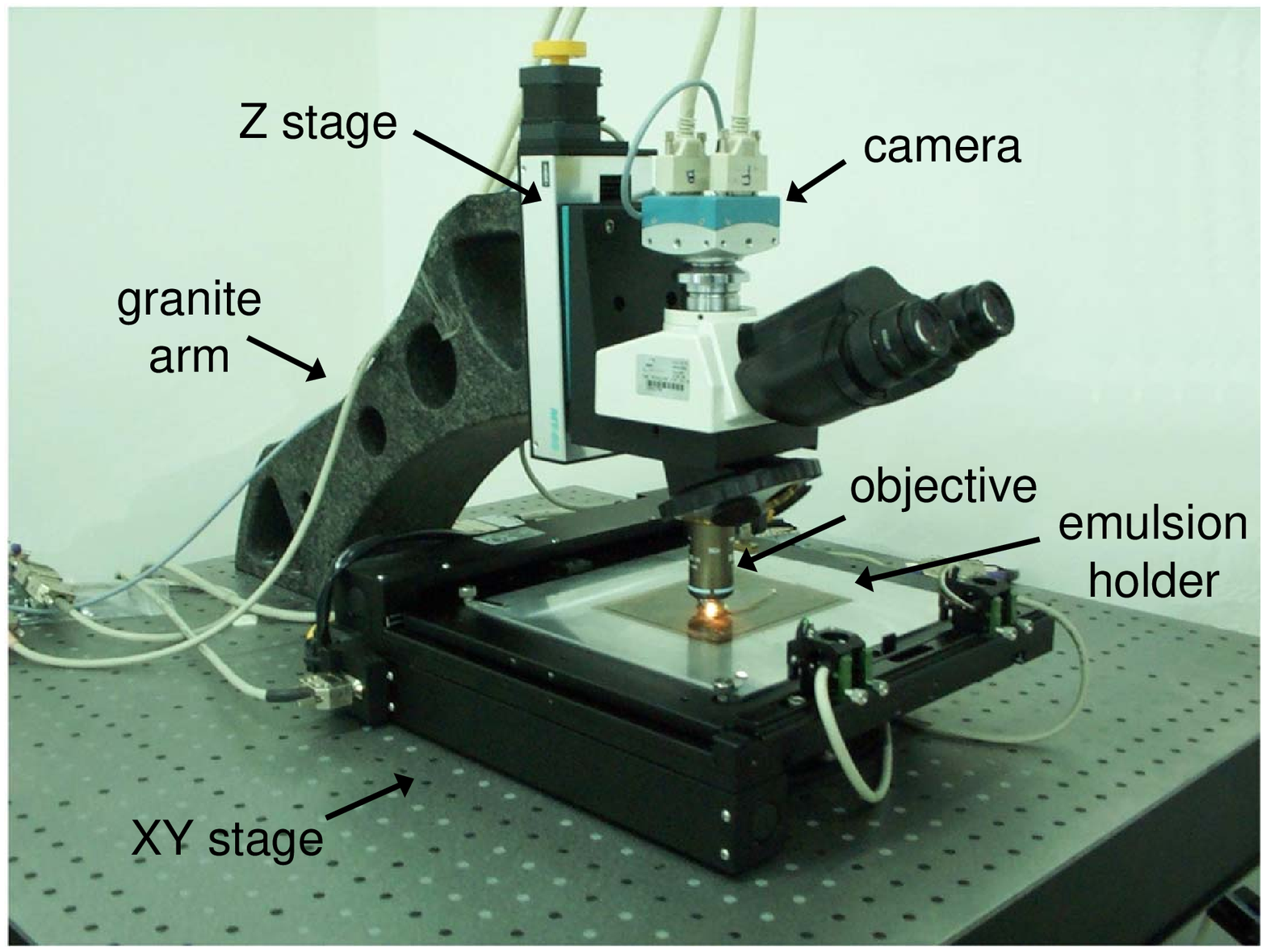,height=5.5cm}
      \epsfig{figure=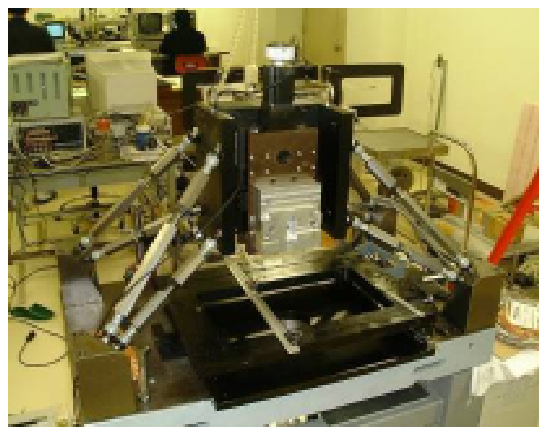,height=5.5cm} }
\caption {Photograph of one of the ESS microscopes (left) and of the
  S-UTS (right).}
\label{fig:ESS}
\end{center}
\end{figure}

{\bf Nuclear emulsions and their scanning.} The basic target module is a 
{\it ``brick''}, consisting 
of a sequence of 56 lead plates (1 mm thick) and 57 emulsion 
layers. A brick has a size of $10.2 \times 12.7$ cm$^2$, a 
depth of 7.5 cm (10 radiation lengths) and a weight of 8.3 kg. Two 
additional emulsion sheets, the {\it changeable 
sheets} (CS), are glued on its downstream face. The bricks are arranged in 
walls. Within a brick, the achieved spatial resolution is $< 1~\mu$m and 
the angular resolution is $\sim$2 mrad.  
 To provide a $\nu$ interaction trigger and to identify the brick in which 
the interaction took place, the brick walls are complemented by walls of 
target trackers and a muon spectrometer \cite{gg}. \par
The bricks 
are made by the {\it Brick Assembling Machine} 
(BAM), which consists of robots for the mechanical packing of the bricks.
 The BAM is  installed in the Gran Sasso lab  
and it may produce $\sim 1$ brick every 1-2 minutes. 
The bricks are handled by the {\it Brick 
Manipulator System} (BMS), made of two robots, each operating at one 
side of the detector. An arm is used to insert the bricks. The extraction 
of a brick, in the region  
indicated by the electronic detectors, is done by a vacuum sucker of the 
BMS.\par 

\begin{figure}[!ht]
\begin{center}
\mbox{\epsfig{figure=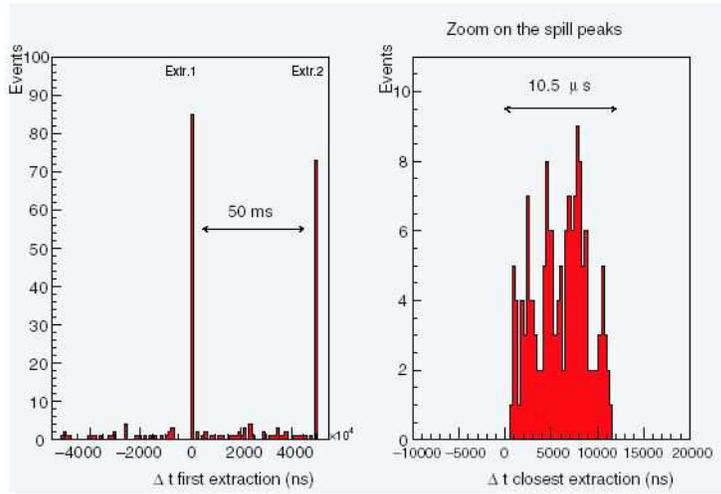,height=7cm}}
\caption {Time distribution of muon events collected by OPERA in the CNGS 
neutrino test 
run. The muon event time difference with respect to the closest extraction is 
shown in the right histogram.}
\label{fig:time}
\end{center}
\end{figure}

A fast automated scanning system is needed to cope with the daily analysis 
of a large number of emulsion sheets. The minimum required scanning speed is 
$\sim20$ cm$^2$/h per emulsion layer ($44 ~\mu$m thick). 
It corresponds to an increase in speed of one order of magnitude 
with respect to past systems \cite{TS,SYSAL}. For this purpose were 
developed the {\it European Scanning System} (ESS) in Europe 
\cite{ESS} and the 
{\it S-UTS} in Japan \cite{SUTS}. \par

\begin{figure}[!ht]
\begin{center}
\mbox{\epsfig{figure=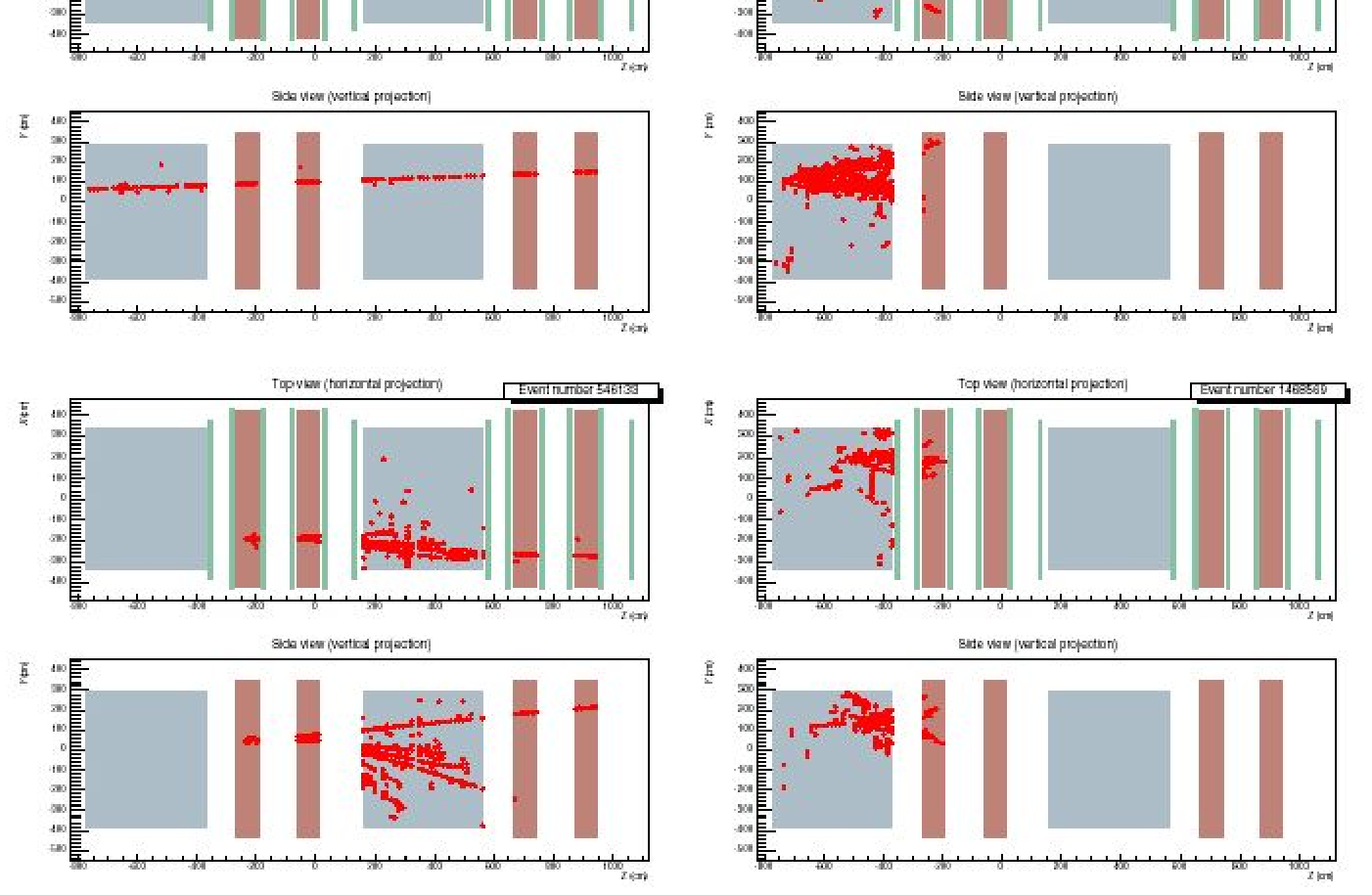,height=10cm}}
\caption {Display of OPERA neutrino muon events from the CNGS test run. 
For each event the top and side views are shown, respectively. 
The targets are indicated in blue, the spectrometers in light brown, Target 
Tracker and RPC hits in red. Top left: $\nu_{\mu}$ CC interaction in the rock 
upstream of the detector; top-right and bottom-right: $\nu_{\mu}$ CC and NC 
interactions in the target material; bottom-left: $\nu_{\mu}$ CC 
interaction in 
the iron of the spectrometer.}
\label{fig:display4}
\end{center}
\end{figure}

The main components of the ESS microscope are shown in Fig. 
\ref{fig:ESS} left: 
 (i) a high quality, rigid and vibration-free support table; (ii) a motor 
driven scanning stage for horizontal (XY) motion; (iii) a granite arm; (iv) 
a motor driven stage mounted vertically (Z) on the granite 
arm for focusing; (v) optics; (vi) digital camera for image grabbing 
mounted on the vertical stage; (vii) 
an illumination system. The emulsion 
is placed on a glass plate (emulsion holder) and its flatness is 
guaranteed by a vacuum system. 
By adjusting the focal plane of the objective, the $44~\mu$m 
emulsion thickness is spanned and 16 tomographic images 
of each field of view are taken at equally spaced depths.
 The images are digitized, converted into a grey scale 
of 256 levels, sent to a vision processor board and analyzed to recognize 
sequences of aligned grains. The three-dimensional structure of a track 
in an emulsion layer ({\it microtrack}) is reconstructed by combining 
clusters belonging to images at different levels. Each microtrack 
pair is connected across the plastic base to form the {\it base track}. A set 
of connected base tracks forms a {\it volume track}.   
The Japanese S-UTS system, Fig. \ref{fig:ESS} right, is based
on hardware designed and made in Nagoya; the software 
system is mounted in specially designed electronic boards.

\section{The first test run with CNGS neutrinos. Conclusions}
A detailed description of the CNGS operation during the August test run may be 
found in \cite{CNGS}. During this run an integrated intensity of 
$7.6 \times 10^{17}$ p.o.t. was delivered. The accuracy in the time 
synchronization between CERN and Gran Sasso was better than 100 ns. The beam 
time spill is shown in Fig. \ref{fig:time}. Event samples recorded by LVD and 
OPERA are shown in Figs. \ref{fig:display} and \ref{fig:display4}, respectively. \par

\begin{figure}[!ht]
\begin{center}
\mbox{\epsfig{figure=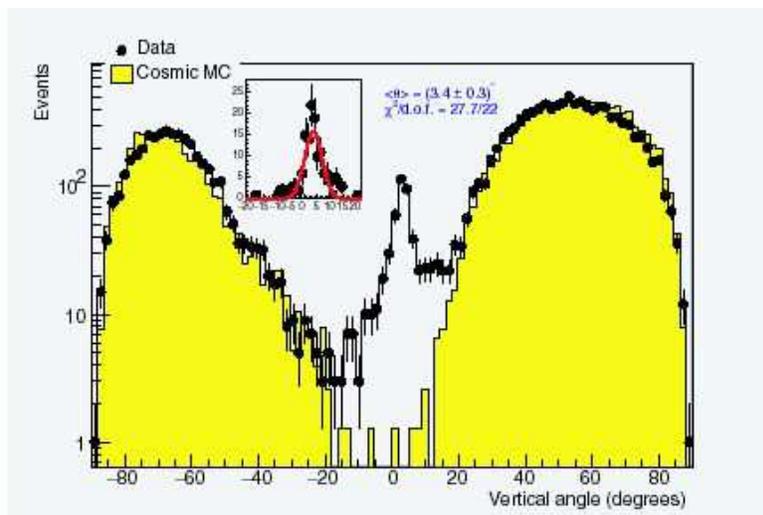,height=7.5cm}}
\caption {Angular distribution of CNGS beam-induced muons and cosmic-muon events taken with the electronic detectors (black points). The histogram is the prediction from cosmic simulations. The inset shows the on-time beam muon events.}
\label{fig:distribution}
\end{center}
\end{figure}

OPERA recorded 319 neutrino events, consistent with the 300 events expected 
on the basis of the delivered integrated intensity. The $\theta$ angular 
distribution with respect to the horizontal axis is shown 
in Fig. \ref{fig:distribution}. In the same figure, the distribution of 
simulated cosmic-ray muons is also shown. In the inset is shown the angular distribution of beam events; a Gaussian fit to the $\theta$ angle distribution 
of these events on-time with the beam yielded a mean muon angle 
of 3.4$^\circ$ in agreement with the value of 3.3$^\circ$, expected for 
neutrinos originating from CERN and travelling under the earth surface to 
the LNGS underground halls. \par

\begin{figure}[!ht]
\begin{center}
{\centering\resizebox*{!}{6cm}{\includegraphics{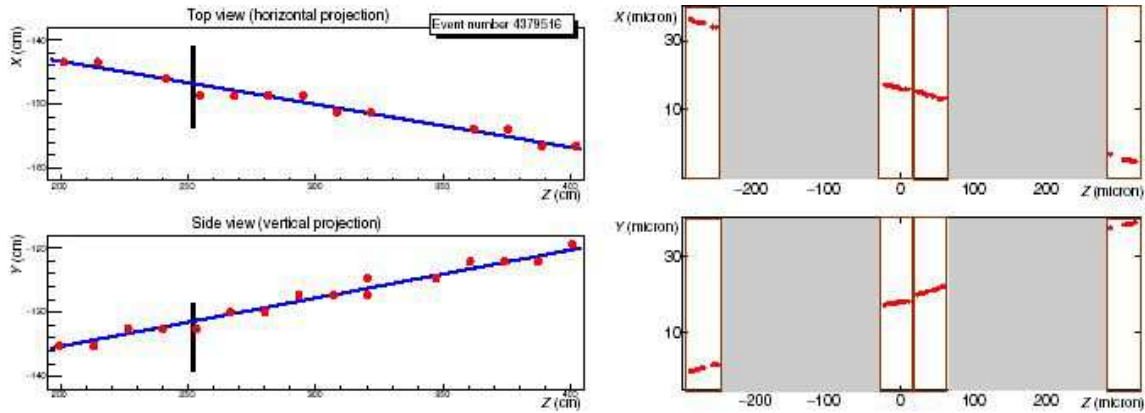}}\par}
\caption {Left: display of one event with the muon passing through electronic 
detectors and (right) the corresponding four microtracks in the 
``changeable sheets'' emulsion doublet.}
\label{fig:1event}
\end{center}
\end{figure}

During the August test run a succesful test of the CS procedure was performed by using an emulsion detector plane consisting of a matrix of $15 \times 20$ 
individual CS doublets inserted in the SM2 target. 9 muons produced by 
neutrino 
interactions in the rock surrounding the detector crossed the CS plane surface;
most of them were found by scanning the emulsion films, see 
Fig \ref{fig:1event}. The test proved the capability in passing from the 
centimetre scale of the electronic tracker resolution to the micrometric 
resolution of nuclear emulsions.\par
In conclusion the CNGS neutrino beam performed well in the first test run. 
And the Borexino, LVD and OPERA experiments at Gran Sasso recorded a number 
of events. \par
In particular OPERA recorded 319 muon events in agreement with expectations. 
The reconstructed zenith-angle distribution from muon tracks in 
centered at 3.4$^\circ$, as expected for neutrinos originating at CERN and 
travelling under the earth surface to LNGS. 
A test of the association between muon tracks reconstructed by electronic 
detectors and with emulsions was successfully performed, proving the 
capability of passing from the centimeter scale of electronic detectors to 
the micron scale of nuclear emulsions.\\

{\normalsize

I would like to acknowledge the cooperation of the CNGS team at CERN, of the 
LNGS staff, and of the Borexino, LVD and OPERA experiments. I thank Ms. 
A. Casoni for typing the manuscript and Dr. M. Giorgini for advice.

}
\end{document}